\documentclass[twocolumn,prl,superscriptaddress,aps,longbibliography,showpacs,preprintnumbers,amsmath,amssymb,floatfix]{revtex4-1}
\usepackage{titlesec}
\usepackage{amsmath}
\usepackage{graphicx}
\usepackage[ansinew]{inputenc}
\usepackage{array}
\usepackage{color}
\usepackage{amsxtra}
\usepackage{amstext}
\usepackage{amssymb}
\usepackage{latexsym}
\usepackage{gensymb}
\usepackage{dsfont}
\usepackage{braket}
\usepackage[caption=false]{subfig}
\usepackage[makeroom]{cancel}

\usepackage{dcolumn}
\usepackage{bm}
\usepackage{bbm}
\usepackage{braket}
\usepackage{mathrsfs}
\usepackage{amstext}
\usepackage{euscript}
\usepackage{txfonts}
\usepackage{soul}

\usepackage[unicode=true,
 bookmarks=false,
 breaklinks=false,pdfborder={0 0 1},backref=false,colorlinks=true,citecolor=blue]
 {hyperref}

\newcommand{\ms}[1]{\mbox{\scriptsize #1}}

\makeatletter
\@ifundefined{textcolor}{}
{%
 \definecolor{BLACK}{gray}{0}
 \definecolor{WHITE}{gray}{1}
 \definecolor{RED}{rgb}{1,0,0}
 \definecolor{GREEN}{rgb}{0,1,0}
 \definecolor{BLUE}{rgb}{0,0,1}
 \definecolor{CYAN}{cmyk}{1,0,0,0}
 \definecolor{MAGENTA}{cmyk}{0,1,0,0}
 \definecolor{YELLOW}{cmyk}{0,0,1,0}
}

\makeatletter
\renewcommand*\env@matrix[1][*\c@MaxMatrixCols c]{%
  \hskip -\arraycolsep
  \let\@ifnextchar\new@ifnextchar
  \array{#1}}
\makeatother

\newcommand{\cref}[1]{Ref.\,\cite{#1}}

\makeatother

\begin{document}


\title{Heisenberg-limited continuous-variable distributed quantum metrology with arbitrary weights}

\date{\today}
\author{Wenchao Ge}
\affiliation{Department of Physics, University of Rhode Island, Kingston, Rhode Island 02881, USA}
\affiliation{Department of Physics, Southern Illinois University, Carbondale, Illinois 62901, USA}

\author{Kurt Jacobs}\affiliation{United States Army Research Laboratory, Adelphi, Maryland 20783, USA}\affiliation{Department of Physics, University of Massachusetts at Boston, Boston, Massachusetts 02125, USA}

\begin{abstract}
Distributed quantum metrology (DQM) enables the estimation of global functions of $d$ distributed parameters beyond the capability of separable sensors. Continuous-variable DQM involves using a linear network with at least one nonclassical input. Here we fully elucidate the structure of linear networks with two non-vacuum inputs which allows us to prove a number of fundamental properties of continuous-variable DQM. While measuring the sum of $d$ parameters at the Heisenberg limit can be achieved with a single non-vacuum input, we show that two inputs, one of which can be classical, is required to measure an \textit{arbitrary} linear combination of $d$ parameters and an arbitrary global function of the parameters. 
We obtain a universal and tight upper bound on the sensitivity of DQM networks with two inputs, and completely characterize the properties of the nonclassical input required to obtain a quantum advantage. This reveals that a wide range of nonclassical states make this possible, including a squeezed vacuum. We also show that for a class of nonclassical inputs local photon number detection will achieve the maximum sensitivity. Finally we show that a general DQM network has two distinct regimes. The first achieves Heisenberg scaling. In the second the nonclassical input is much weaker than the coherent input, nevertheless providing a multiplicative enhancement to the otherwise classical sensitivity.


\end{abstract}

\maketitle
\textit{Introduction}--Quantum metrology is the science of measurements that exploit quantum-mechanical properties to improve sensitivity over classical limits~\cite{giovannetti2006quantum}. 
Multi-parameter quantum metrology~\cite{Humphreys:2013ab, gagatsos2016gaussian,  Polino:19, szczykulska2016multi} broadens the scope of this field with quantum-enhanced sensitivity for estimating simultaneously multiple parameters, paving the way for near-term real-world quantum information applications~\cite{Ang17, ALBARELLI20}.

An emerging direction of multi-parameter quantum metrology is to estimate global properties of the output from multiple sensors that are part of an entangled network prepared with distributed entangled states. This is termed \emph{distributed quantum metrology} (DQM) \cite{Zach:18, GePRL2018, Zhuang:18, Proctor2018PRL}, and has applications to global-scale clock synchronization~\cite{komar2014quantum}, real-time noise characterization~\cite{Majumder:20}, and ultrasensitive positioning~\cite{YiPRL2020, Sun:22}. The global properties of a quantum sensor network could in general be arbitrary analytic functions~\cite{Qian:19, Rubio:20, Qian:21, Bringewatt:21, Ehrenberg:23} of the local parameters at each node. A key component of measuring such functions is the ability to measure an \textit{arbitrary} linear combination of the local parameters~\cite{Qian:19}. Schemes that can realize this at the Heisenberg limit (HL)~\cite{giovannetti2004quantum, zwierz2010general}, i.e., the sensitivity scales as $1/N$ with $N$ the average total number of particles, have been devised using GHZ states of qubits~\cite{Zach:18, Proctor2018PRL} and entangled states generated by a twin-Fock state at a passive linear interferometer~\cite{GePRL2018}. However, it remains challenging to generate scalable discrete-variable multi-partite entanglement~\cite{Yang:24} or higher Fock states for distributed quantum metrology despite recent progress~\cite{Tiedau:19, Uria:20, Rivera:23, deng2023heisenberg, Bell:13, Hong:21, Liu:21, Zhao:21, Valeri:23, Kim:24}. 

Continuous-variable (CV) quantum states~\cite{Braunstein:05}, on the other hand, have been explored for DQM of phase sensing~\cite{GattoPRA19, Guo:20} and displacement sensing~\cite{Zhuang:18, YiPRL2020} due to their experimental scalability~\cite{Loock:00,Yukawa:08, Vogel:14, Zhuang:18} in linear optical networks, where large optical interferometers have already been demonstrated with a couple of hundred modes~\cite{Zhong20,Madsen:2022wo}. Single-mode squeezed vacuum states have been extensively studied among other Gaussian states for distributed phase sensing~\cite{OhPRR20,GattoPRA19,Zhuang:18, Guo:20, OhPRR20}. The first proof-of-principle demonstration of quantum-enhanced sensing of an averaged phase shift used a single-mode squeezed vacuum state distributed among four nodes~\cite{Guo:20}. 

Recently Malitesta, Smerzi, and Pezz\`{e}~\cite{Malitesata23} presented the first CV DQM scheme that can measure arbitrary combinations of the parameters at the Heisenberg limit. This scheme distributes a squeezed vacuum input via a linear network to $d$ separated Mach-Zehnder interferometers (MZIs) that employ $d$ local coherent inputs.  
 
Here we are able to elucidate the full structure of all DQM schemes with an arbitrary linear network and two non-zero inputs. This allows us to prove a number of fundamental results regarding such  networks. First we show that two non-zero inputs, with only one being nonclassical, are the minimum required for Heisenberg-limited measurement of an arbitrary linear combination of the distributed parameters. This result shows that the scheme of Malitesta, Smerzi, and Pezz\`{e}~\cite{Malitesata23} is the simplest possible from the point of view of the inputs \footnote{Since the classical input does not distribute entanglement it can either be distributed locally from a single input, simpler from the point of view of maintaining a single phase reference, or be replaced by separate classical inputs that are local to the parameters being measured.}. 
Second, we show that the error for estimating a global parameter with a linear network with two inputs is universally lower-bounded by $1/\sqrt{N+2n_2\mathcal{W}}$, where $n_2$ is the mean photon number of the coherent-state input and $\mathcal{W}$ is the \textit{metrological power} of the nonclassical state for quadrature sensing. The metrological power was introduced in~\cite{Ge:20Opearational, Ge:23}, scales at most linearly with the energy of the state, and has a simple form for pure states which we give below. We stress that this lower bound on the sensitivity holds for both pure and mixed input states.   

Third, we determine the precise properties required by the non-classical input to obtain a quantum advantage, which involves the first, second, and third moments of the annihilation and creation operators. This shows that a wide range of inputs will realize a quantum advantage. Fourth, we demonstrate that for a fixed non-classical input state the network can be configured to estimate any global function of the parameters with the same sensitivity by tuning the transmissivities of the beam-splitters in the linear network. Fifth, we show that for a class of non-classical inputs photon-number detection performed \emph{locally} at the distributed nodes is sufficient to obtain the maximum sensitivity allowed by the input states. Finally, we show that there are two distinct regimes in which a linear network can be operated. When the energies of both the classical and nonclassical input states are sufficiently large and comparable, the optimal sensitivity can scale as $1/N$, reaching the HL for a class of nonclassical states~\cite{GePRA20}, including squeezed vacuum states (SV) and cat states (CS)~\cite{Our:07, Lewenstein:21}. In the regime in which the nonclassical state is weak compared to the coherent input, the sensitivity is amplified over that with coherent states alone by a factor of the metrological power~\cite{Ge:20Opearational, Ge:23} of the nonclassical state.

 \begin{figure}[t]
\leavevmode\includegraphics[width = 0.9\columnwidth]{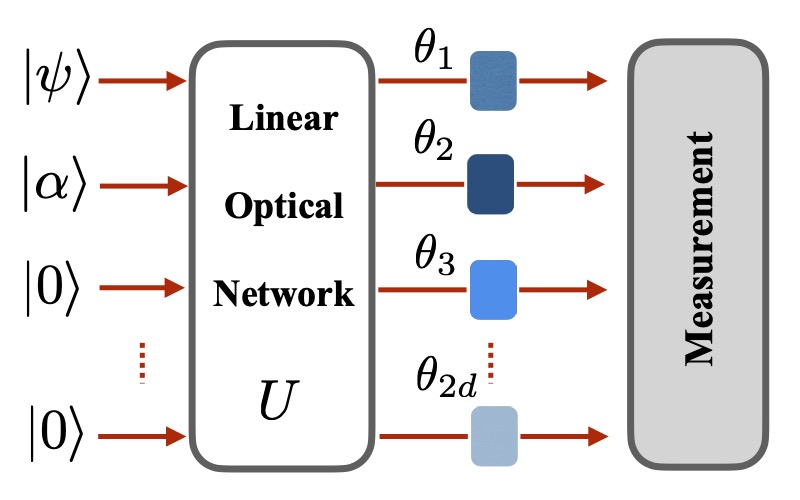}
\caption{Continuous-variable distributed quantum metrology protocol. An input of $\ket{\Psi}=\ket{\psi}\otimes\ket{\alpha}\otimes\ket{0}\cdots\otimes\ket{0}$ is injected to a linear optical network $U$ followed by encoding of the parameters $\bm{\theta}=\left(\theta_1,\theta_2,\cdots,\theta_{2d}\right)$. The measurement step is global in general, but we show local photon-number detection can saturate the lower bound of Eq~\eqref{eq:globalLB} for some inputs.} 
\label{fig1} 
\end{figure} 





\textit{The Model.---} We consider the general distributed quantum metrology scheme~\cite{GePRL2018} for estimating phase shifts in a multiport linear-optical interferometer with one single-mode nonclassical state $\ket{\psi}$ and a coherent state $\ket{\alpha}$ as shown in Fig.~\ref{fig1}. The generalization to a nonclassical mixed state input $\rho$ with $\ket{\alpha}$ is discucssed in the Supplemental Materials~\cite{SM}. To properly evaluate the sensitivity for phase estimation each phase being estimated must be measured with respect to a reference phase contained in a separate mode~\cite{Jarzyna:12}. To estimate $d$ phases we use a separate reference for each mode for a total of $2d$ modes~\footnote{An equivalent scheme of $d+1$ modes with only one reference mode is shown in the Supplemental Materials~\cite{SM}.}. 

The initial product state $\ket{\Psi}=\ket{\psi}\otimes\ket{\alpha}\otimes\ket{0}^{\otimes(2d-2)}$ can generate distributed entangled states in the linear network for the sensing task.  After the network acts on the $2d$ modes the state is given by $|\Psi_{U}\rangle=U\ket{\Psi}$. We denote the annihilation operators for the 2d input (output) modes before (after) the unitary matrix $U$ by $\hat{a}_j (\hat{b}_j)$. The operators are thus related by the network as $\hat{b}^{\dagger }_j=\sum_kU_{jk}\hat{a}^{\dagger}_k$. With the introduction of the phase shifts $\theta_j$, $\ket{\Psi_U}$ is mapped to $\exp(-i\hat{H}(\bm{\theta}))\ket{\Psi_{U}}\equiv \ket{\Psi_U(\bm{\theta})}$, where $\hat{H}(\bm{\theta})=\sum_{j}\theta_j\hat{n}_j$ with $\hat{n}_j=\hat{b}^{\dagger}_j\hat{b}_j$. For simplicity, all sums run from $1$ to $2d$ throughout this manuscript unless otherwise specified.

We aim to estimate a global quantity $ q =2\sum_{j}w_j\theta_j$ by making measurements on $\ket{\Psi_U(\bm{\theta})}$, preceded by an additional linear-optical unitary $V$ if necessary (Fig.~\ref{fig1}). Here $w_{2i-1}=-w_{2i}$ such that each phase is paired with its reference: $ q =2\sum_{i=1}^dw_{2i}\left(\theta_{2i}-\theta_{2i-1}\right)$, and the 1-norm $||\bm{w}||_1\equiv\sum_{k}|w_k|$ is normalized to unity. The factor $2$ in $q$ is to ensure that $q$ is the average of the phase differences $\theta_{2i}-\theta_{2i-1}$ when $w_{2i}=1/(2d)$ $(i=1,2,\cdots, d)$. Our primary tool is the multi-parameter quantum Cram\'er-Rao bound~\cite{Liu_2020}, which states that a set of unbiased estimators $\Theta_j$ for parameters $\theta_j$ satisfy
\begin{align}
\label{eq:qcrb}
{\rm Cov}(\bm{\Theta})\geq \left(M\mathcal{F}\right)^{-1}.
\end{align}
Here $M$ is the number of repetitions~\cite{kay1993fundamentals} which is taken to be one in the following analysis for the asymptotic per-trial limit, and the covariance matrix is defined by its matrix elements as ${\rm Cov}(\bm{\Theta})_{jk}\equiv E[(\Theta_j-\theta_j)(\Theta_k-\theta_k)]$, where $E[X]$ is the expected value of the quantity $X$, and in this context the quantum Fisher information matrix (QFIM) $\mathcal{F}$ is defined by its matrix elements as
$\mathcal{F}_{jk}\equiv 2\bra{\Psi_U}\{\hat{n}_j,\hat{n}_k\}\ket{\Psi_U}  - 4\bra{\Psi_U}\hat{n}_j\ket{\Psi_U}\bra{\Psi_U}\hat{n}_k\ket{\Psi_U}.$
Using $Q=2\sum_{j}w_j\Theta_j$ as an unbiased estimator of $ q $, the uncertainty $\Delta^2q\equiv E[(Q-q)^2]=4\sum_{j,k}w_j{\rm Cov}(\bm{\Theta})_{jk}w_k$, as has been shown in Ref.~\cite{GePRL2018},
\begin{align}
\label{eq:q_bound}
\Delta^2 q\geq 4\boldsymbol{w}^T\mathcal{F}^{-1}\boldsymbol{w}.
\end{align}
It has been shown that only input states that concentrate the non-classical resources in a relatively small number of the input modes can achieve the HL for all the modes~\cite{GePRL2018}.



\textit{Quantum Fisher Information Matrix.---} The key element in determining the global estimation sensitivity in Eq.~\eqref{eq:q_bound} is the QFIM. We are able to write the QFIM for distributed phase sensing in a linear network with two initially separable input states as
\begin{align} \label{eq:QFIM} \mathcal{F}=c_u\boldsymbol{u}\boldsymbol{u}^T+c_v\boldsymbol{v}\boldsymbol{v}^T+c_s \left[ \boldsymbol{u}\boldsymbol{v}^T+\boldsymbol{v}\boldsymbol{u}^T \right] +\mathcal{N},
\end{align}
where $c_u=4(\nu_1-n_1)$, $c_v=8n_2\mathcal{W}$, and $c_s=8\Re[(\beta_1+\alpha_1/2-n_1\alpha_1)\alpha^{\ast}]$ with $\mathcal{W} = (n_1-|\alpha_1|^2+|\xi_1-\alpha_1^2|)$, $\nu_1=\braket{\psi|(a_1^{\dagger}a_1)^2|\psi}-n_1^2$, $\beta_1=\braket{\psi|a^{\dagger}_1a_1a_1|\psi}$, $n_1=\braket{\psi|a_1^{\dagger}a_1|\psi}$, $\xi_1=\braket{\psi|a_1^2|\psi}$, and $\alpha_1=\braket{\psi|a_1|\psi}$. Here $\mathcal{W}$ is the metrological power of the single-mode quantum state $\ket{\psi}$ for quadrature measurement~\cite{Ge:23, Ge:20Opearational}, which quantifies its ability for quantum-enhanced sensing compared to coherent states. The vectors $\boldsymbol{u}=(|U_{11}|^2,\cdots, |U_{2d, 1}|^2)^T$ and $\boldsymbol{v}=\left(U_{11}U_{12}^{\ast},\cdots,U_{2d,1}U_{2d,2}^{\ast}\right)^T$ are determined by the linear network in transforming from the two input modes into the $2d$ output modes. We have optimized the phases of these elements and that of the coherent state for the QFIM such that $\boldsymbol{v}$ is real and $\arg[\alpha^2]=\arg[\xi_1-\alpha_1^2]$. The diagonal matrix $\mathcal{N}=4\text{Diag}\{|U_{11}|^2n_1+|U_{12}|^2n_2,\cdots,|U_{2d,1}|^2n_1+|U_{2d,2}|^2n_2\}$ describes the distribution of the total input state energy in the linear network, which contributes to the standard quantum limit (SQL) in distributed phase sensing, i.e., the scaling of $1/N^{1/2}$ with $N=n_1+n_2$, while the terms $c_u$ and $c_v$ are quantum-enhanced contributions due to the variance property of the nonclassical state and the distributed entangled state generated in the network, respectively. For a mixed nonclassical state $\rho$ and a coherent state at the input port, $c_u, c_v, c_s$ and $\mathcal{N}$ in Eq.~\eqref{eq:QFIM} will be generalized with the corresponding values of $\rho$~\cite{SM}. 

\emph{Single-mode input.---}We now show that with only one non-vacuum input mode, the distributed quantum metrology scheme of the general form in Fig.\ref{fig1} cannot estimate a linear combination of phases containing both positive and negative weights with Heisenberg-limited sensitivity. The corresponding QFIM with $\alpha=0$ reduces to $\mathcal{F}=c_u\boldsymbol{u}\boldsymbol{u}^T+\mathcal{N}$. If we neglect the contribution from the second term, the setup may achieve quantum-enhanced sensitivity only when the weighting vector $\boldsymbol{w}$ is parallel to $\boldsymbol{u}$, whose elements are positive-definite. This reveals that the sensitivity enhancement of a single-mode quantum state relies on the amplified quadrature variance of the state, which can only translate to positive weights in collective phase estimation (e.g.~\cite{Zhuang:18,Guo:20, OhPRR20, GattoPRA19}). We also observe from Eq.~\eqref{eq:QFIM} that some single-mode nonclassical states with low photon-number variance, such as Fock states, cannot even beat the SQL when injected into a linear network alone~\cite{Takeoka17}.

\emph{Quantum-enhanced distributed phase sensing.---}We now show that Heisenberg-limited sensing of a linear combination of parameters with arbitrary weightings can be obtained with one coherent and one nonclassical input. According to Eq.~\eqref{eq:q_bound}, the optimal sensitivity for a given input state corresponds to a particular weighting $\boldsymbol{w}$ that is parallel to the eigenvector of the largest eigenvalue of $\mathcal{F}$. For quantum enhanced sensing either or both of $c_u, c_v$ must be much greater than $N$ for large $N$. Thus $\boldsymbol{w}$ has to lie in the plane supported by $\boldsymbol{u}$ and $\boldsymbol{v}$. Due to the condition $\sum_jw_j=0$, the weighting vector $\boldsymbol{w}$ can only be chosen to be parallel to $\boldsymbol{v}$. Otherwise, the contribution would be dominated by $O(1/N)$ in the estimation variance~\cite{SM}. We show that for arbitrary weights~\cite{SM} 
\begin{align}
\label{eq:globalLB}
   \Delta q\geq \frac{1}{\sqrt{N+\frac{c_v}{4}-\frac{c_s^2}{16N+4c_u}}} \ge \frac{1}{\sqrt{N+2n_2\mathcal{W}}}.
\end{align}
This relation holds for an arbitrary state $\rho$ mixing with a coherent state at the input. The first lower bound is obtained by choosing the unitary elements of the linear network $U_{j1}=\sqrt{|w_j|}$ and $U_{j2}=w_j/\sqrt{|w_j|}$ such that $\boldsymbol{v}=\boldsymbol{w}$. The second lower bound is saturated for those nonclassical states with vanishing $c_s$, including squeezed vacuum states, squeezed thermal (ST) states,  cat states, and Fock states~\cite{Ge:20Opearational}. These results of the multiport interferometry reduce to that of the MZI injected with one coherent state and one nonclassical mode~\cite{pezze2008mach, Lang:13} when we take $d=1$. For a fixed $N$, the quantum-enhanced sensitivity depends on the ratio between the mean photon numbers of the two input modes and the metrological power of the state~\cite{Ge:20Opearational}, $\mathcal{W}$, where the latter is zero for classical states (including coherent states) and is maximal for the squeezed vacuum state for a given value of $n_1$ ~\cite{Ge:20Opearational, Ge:23}. We show that for a fixed $N$, the optimal quantum state at the first port in the distributed quantum phase sensing with a coherent state at the second input port is a squeezed vacuum state, generalizing the case of the two-mode MZI~\cite{Lang:13}. 

\begin{figure}
\centering
\subfloat{\includegraphics[width=.8\columnwidth]{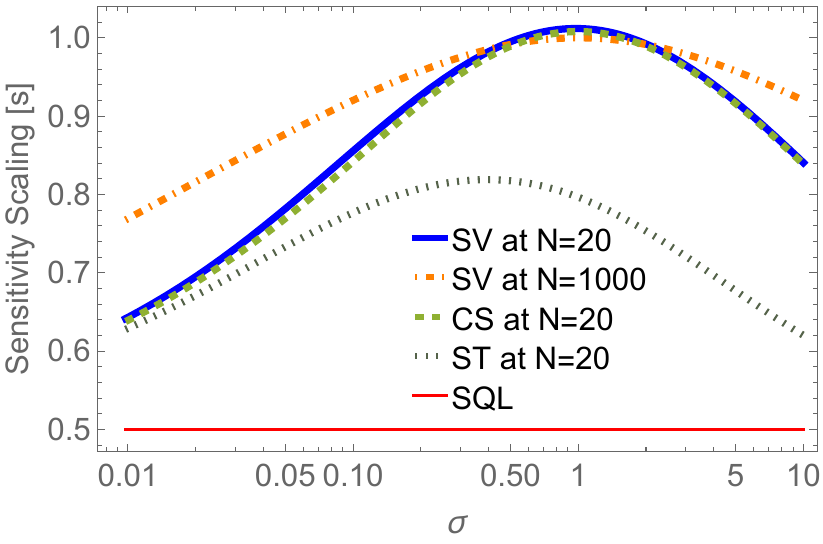}}
\\[1ex]
\subfloat{\includegraphics[width=.8\columnwidth]{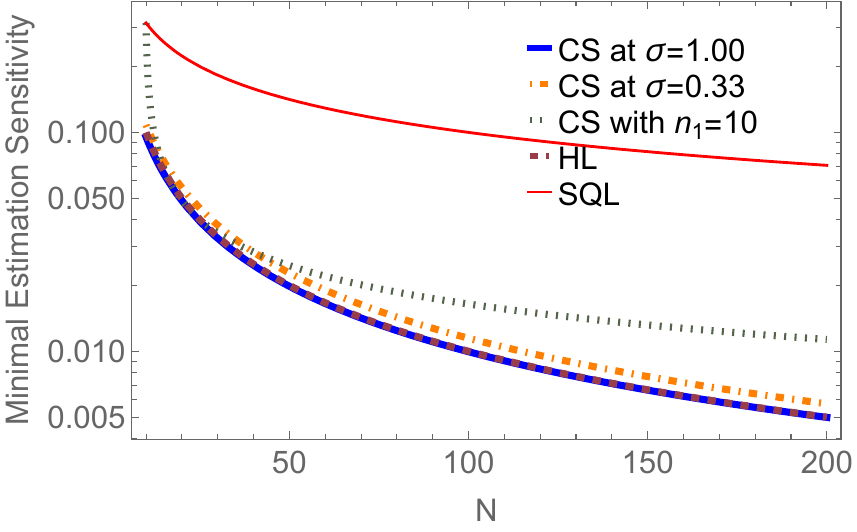}}
 \caption[]{(Color online). (a) The sensitivity scaling, $s$, as a power of $N$ vs the ratio $\sigma$ of the input mean photon numbers for squeezed vacuum (SV) states, a cat state (CS), and a squeezed thermal (ST) state~\cite{SM}. (b) The minimal estimation sensitivity for different cat states as a function of $N$.}
\label{fig:ff}
\end{figure}

For a fixed $N\gg1$, the scheme can achieve the HL for distributed phase sensing with arbitrary weights by requiring $\mathcal{W}\sim n_2\sim N/2$. 
On the other hand, when the nonclassical state is much weaker than the coherent input (that is, $1<\mathcal{W}\ll n_2$)~\cite{Ge:23}, the scheme is useful for amplifying the precision of distributed sensing that employs relatively strong classical light with an amplification factor of $1/\sqrt{2\mathcal{W}}$. The latter is of particular interest for the state-of-the-art experiments since even the most energetic nonclassical states can be outperformed by weak lasers~\cite{Lang:13, Meylahn:22}. By injecting one coherent state into the network, we include the reference energies when evaluating the scaling of the sensitivity. In the top panel of Fig.~\ref{fig:ff}, we plot the curve of the sensitivity scaling ($s\equiv \log_N\sqrt{N+2n_2\mathcal{W}}$) as a function of the input photon ratio $\sigma\equiv n_1/n_2$. We show that for SV and CS, the optimal scaling appears at $\sigma=1$, which is due to the fact that $\mathcal{W}$ is a linear function of $n_1$ for these states~\cite{GePRA20}. Otherwise, the optimal scaling ratio shifts to other values depending on the relation between $\mathcal{W}$ and $n_1$, such as the ST state. In the lower panel, we plot curves of the minimal estimation sensitivity of a global parameter $q$ vs the total input photon number $N$ for different input states at different photon ratios. We observe that the minimal sensitivity is robust against changes of the photon ratio near the optimal ratio, and that the sensitivity is enhanced compared to the SQL using a CS with a fixed mean photon number $n_1=10$ when $N\gg n_1$. 


\textit{Estimation of arbitrary functions.---}Our quantum sensor network can be employed to estimate arbitrary analytic functions of local parameters with quantum enhancement beyond separable sensors. Function estimation has been analyzed using discrete variables of qubits or photonic GHZ states~\cite{Qian:19} to show Heisenberg scaling with the number of parameters. Here we apply the protocol to our DQM scheme with continuous-variable separable input states.

The optimal procedure for estimating a function of multiple parameters can be achieved in two steps~\cite{Qian:19}. First, we make an unbiased estimate $\tilde{\bm{\theta}}$ of $\bm{\theta}$ using a small portion, $N_1$, of the total energy. Then, we make an unbiased measurement $\tilde{q}$ of the quantity  $q=\nabla f(\tilde{\bm\theta})\cdot(\bm{\theta}-\tilde{\bm\theta})$ based on our initial estimates $\tilde{\bm{\theta}}$ using the rest of the total energy $N_2$. Here $q$ is the first-order expansion of a given function $f(\bm{\theta})$ at $\tilde{\bm{\theta}}$. The estimation of $q$ can be done efficiently using the distributed quantum sensing protocol by taking the weight vector $\bm{w}$ to be parallel to $\nabla f(\tilde{\bm\theta})$, noting that $\theta_{2j}$ and $\theta_{2j-1}$ appears as a difference $\theta_{2j}-\theta_{2j-1}$ in $f(\bm{\theta})$ due to the DQM protocol. Clearly, to estimate an arbitrary function, the quantum sensor network must be able to estimate an arbitrary linear combination of the parameters $\bm{\theta}$. According to Ref.~\cite{Qian:19}, the sensitivity employing this two-step protocol is
\begin{align}
\Delta^2f(\bm\theta)\ge \mbox{V}[\tilde{q}]+\sum_{i,j}\frac{2f^2_{ij}(\bm\theta)+f_{ii}(\bm\theta)f_{jj}(\bm\theta)}{4}\text{V}[\tilde{\theta}_i]\text{V}[\tilde{\theta}_j], 
\end{align}
where $f_{ij}(\bm\theta)\equiv\partial^2f(\bm\theta)/\partial\theta_i\partial\theta_j$ and $\text{V}[\tilde{x}]\equiv E[(\tilde{x}-x)^2]$ for any variable $x$. Here the first term on the RHS of the inequality is the error from the global estimation stage while the second term describes any residual error, up to $O((\tilde{\bm{\theta}}-\bm{\theta})^4)$, from the first step 
of the protocol, after it has been corrected by the 
measurement of the linear combination. By choosing $\bm{w}\ // \ \nabla f(\tilde{\bm\theta})$, the sensitivity of the first term using our network is given by $ ||\nabla f(\tilde{\bm\theta})||_1^2/(4N_2^{2s})$, where the factor is due to the normalization of $\nabla f(\tilde{\bm\theta})$. The second term has a much smaller scaling for a large total number of photons $N$. Under the optimal energy allocation~\cite{Qian:19} when $N_2=O(N)$ and $N_1=O(N^{\frac{2s+1}{4s+1}})$, the sensitivity scales as
\begin{align}
    \Delta^2f(\bm\theta)\ge \frac{||\nabla f(\tilde{\bm\theta})||_1^2}{4N^{2s}}=\frac{||\nabla f(\tilde{\bm\theta})||_1^2}{4N+8n_2\mathcal{W}}.
\end{align}
For sensing at the Heisenberg limit ($s=1$), the optimal resource allocation reduces to $N_1=O(N^{3/5})$ and $N_2=O(N)$.


 \begin{figure}[t]
\leavevmode\includegraphics[width = 0.7\columnwidth]{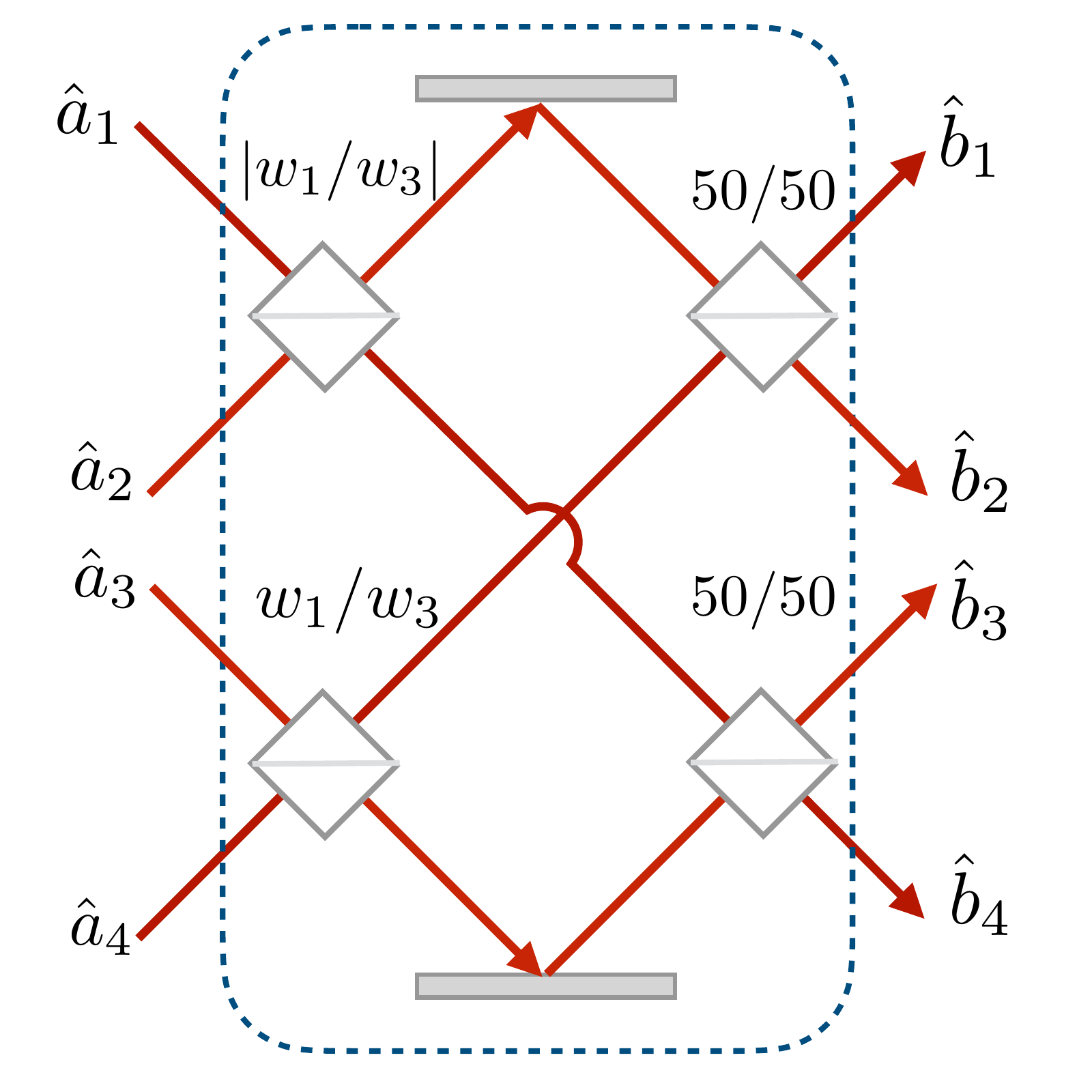}
\caption{A four-mode linear network $U$ for generating useful entangled states for continuous-variable distributed quantum metrology. The ratio of the transmittance to the reflectance of each beam splitter is denoted.   } 
\label{fig:4-mode} 
\end{figure} 


\textit{Experimental realizations.---}The protocol we have presented is readily accessible with current experimental technologies. The requirements of our protocol include a coherent laser resource, single-mode nonclassical light, beamsplitters and phase shifters, and local photon detection measurements~\cite{Daryl:04,kardynal2008avalanche,Photon-statistics:19, Madsen:2022wo, Birrittella:21, Changhun:17}. For the nonclassical input, a squeezed vacuum or a cat state is sufficient, which have been realized in many experiments~\cite{YuPRA:20, Our:07,Guo:20, Lewenstein:21, Meylahn:22, Park:24}.  
For the linear optical network, we present a four-mode system as an example in Fig.~\ref{fig:4-mode}. The output modes are related to the input modes as $\hat{b}^{\dagger}_j=\sum_k U_{jk}\hat{a}^{\dagger}_k$, where the unitary transformation is given by
\begin{gather}
 U = \begin{bmatrix} \sqrt{|w_1|}& \frac{w_1}{\sqrt{|w_1|}} & \sqrt{|w_3|} & \frac{w_3}{\sqrt{|w_3|}}  \\  \sqrt{|w_1|}& -\frac{w_1}{\sqrt{|w_1|}} & \sqrt{|w_3|} & -\frac{w_3}{\sqrt{|w_3|}} \\ \sqrt{|w_3|}& \frac{w_3}{\sqrt{|w_3|}} & -\sqrt{|w_1|} & -\frac{w_1}{\sqrt{|w_1|}} \\ \sqrt{|w_3|}& -\frac{w_3}{\sqrt{|w_3|}} & -\sqrt{|w_1|} & \frac{w_1}{\sqrt{|w_1|}} \end{bmatrix}.
\end{gather}
As seen in Fig.~\ref{fig:4-mode}, any linear combination of the two phase differences can be measured merely by tuning the transmittances of the first two beam splitters. 

To achieve the ultimate lower bound of Eq.~\eqref{eq:globalLB}, we need to determine the optimal measurement to make on the output modes. Inspired by the Mach-Zehnder interferometry~\cite{Lang:13}, where photon-number-resolving detection is an optimal measurement when a coherent input is paired with certain nonclassical states, we prove that in our scheme photon-number-resolving detection that is local to each parameter will maximize the quantum Fisher information for some nonclassical input states. In the Supplemental Materials~\cite{SM}, we prove that when 
$\bra{\bm {n}^D}e^{i\sum_{l}\hat{n}_l^{W}q_l}U\ket{\Psi}$ is real up to a trivial global phase, and $\alpha_1=\beta_1=0$, the classical Fisher information from photon-number detection can saturate the quantum Fisher information. Here $\bra{\bm {n}^D}$ is a photon number detection result $\bm {n}^D$ for the $2d$ modes. While photon-number-resolving detection remains challenging, recent developments~\cite{Daryl:04,kardynal2008avalanche,Photon-statistics:19, Madsen:2022wo} show promise in this direction. The HL DQM scheme also motivates exploration of other measurements, such as parity measurement~\cite{Birrittella:21}, and homodyne detection~\cite{Changhun:17, Hou:24} to extract the metrological gain from the nonclassical inputs. 

\emph{Conclusion.---}Here we have been able to fully elucidate the structure of the quantum Fisher information for distributed phase measurement in a linear network with one classical and one nonclassical input. This has allowed us to show that for arbitrary distributed quantum sensing at the Heisenberg limit two inputs are the minimum required, and to identify precisely the properties required of the nonclassical input. We have shown how our network can be used for the Heisenberg-limited measurement of arbitrary global functions of the local phase shifts. In the problem of obtaining Heisenberg scaling it is the input states that have to-date received the most recent attention. Nevertheless, they are only half of the problem. One must also find the most practical measurements that will allow achieving the HL. While this question remains largely open, we have shown here that local photon-number resolving detection of the output modes will achieve the HL for certain non-classical input states. In showing that many kinds of nonclassical input can be used to realize arbitrary distributed sensing at the HL we have opened up a new line of research to determine which kinds of measurements will realize the HL with which classes of nonclassical inputs. Our elucidation of the structure of the QFI for distributed sensing has provided a powerful tool in this endeavor. 

\textit{Note added.} A recent related work appeared~\cite{pezze2024distributed}, which presented an optimal local-measurement scheme for distributed quantum phase sensing with an array of $d$ spatially-distributed MZIs using a generic nonclassical pure state and $d$ local coherent states.

\section{Acknowledgements}
The research of W.G. is supported by NSF Award 2243591.



\bibliography{DQMHL}

\begin{widetext}
\title{Supplementary Materials}
\maketitle

\newpage
\appendix\newpage\markboth{Appendix}{}
\setcounter{equation}{0}
\renewcommand{\thesection}{S}
\numberwithin{equation}{section}

\section{Supplemental Materials}
\section{S1. The lower bound on $\boldsymbol{w}^T\mathcal{F}^{-1}\boldsymbol{w}$ without $\mathcal{N}$}
\subsection{S1.1. Decomposition of the weighting vector}
In the limit of quantum-enhanced metrology, we neglect $\mathcal{N}$ in Eq.~(3) in the main text. For $\text{Dim}[\mathcal{F}]>2$, $\mathcal{F}$ is singular and so the RHS of Eq.~(3) will be  infinite unless $\boldsymbol{w}$ lies within the space spanned by $\boldsymbol{u}$ and $\boldsymbol{v}$. 
In general, we can thus assume $\boldsymbol{w}=g\left(\cos\alpha \boldsymbol{u}+\sin\alpha \boldsymbol{v}\right)$. We now evaluate the sum of the vector component on both sides of this equation. Note that the given the form of the elements of $\boldsymbol{v}$, the sum of the elements of $\boldsymbol{v}$ is the inner product of two columns of $U$ and is therefore zero. The sum of the elements of $\boldsymbol{w}$ is also zero (due to the need for each measured phase to have a phase reference). Summing both sides of the above equation for $\boldsymbol{w}$ therefore gives $0=g\cos\alpha$. We can conclude the multimode interferometer can only be used to estimate a global phase $q=\boldsymbol{w}\cdot \boldsymbol{\theta}$ with a proper phase reference when $\boldsymbol{w}=g\boldsymbol{v}$ in the limiting case when $\mathcal{N}$ is neglected. Since $||\boldsymbol{v}||_1=\sum_{j}|U_{j1}U_{j2}^{\ast}|\le 1$ and $||\boldsymbol{w}||_1=1$, we require $|g|\ge1$.

\subsection{S1.2. Effective quantum Fisher information}
Neglecting $\mathcal{N}$ for large $N$, we have $\mathcal{F}\approx c_u \boldsymbol{u}\boldsymbol{u}^T+c_v\boldsymbol{v}\boldsymbol{v}^T+c_s\boldsymbol{u}\boldsymbol{v}^T+c_s\boldsymbol{v}\boldsymbol{u}^T$. To simplify the analysis, we first assume input states for which $c_s=0$ and after completing that analysis we will see how it is modified by including the cross terms.

Now, we define two orthonormal vectors $\boldsymbol{s}_1$ and $\boldsymbol{s}_2$ such that $\boldsymbol{u}=\sqrt{N_u}\boldsymbol{s}_1$ and $\boldsymbol{v}=\sqrt{N_v}(\cos\theta \boldsymbol{s}_1+\sin\theta\boldsymbol{s}_2)$, where $N_u, N_v$ are the normalization factors of the vectors $\boldsymbol{u}$ and $\boldsymbol{v}$. So we can write $\mathcal{F}$ as 
\begin{align}
    \mathcal{F} = \left( \begin{array}{cc}
      c_u N_u + c_v N_v \cos^2\theta  & c_v N_v \cos\theta\sin\theta  \\
       c_v N_v \cos\theta\sin\theta    & c_v N_v \sin^2\theta
    \end{array}  \right). 
\end{align}
The inverse of $\mathcal{F}$ is then 
\begin{align}
    \mathcal{F}^{-1} & = \frac{1}{  (c_u N_u + c_v N_v \cos^2\theta) c_v N_v \sin^2\theta - c_v^2 N_v^2 \cos^2\theta\sin^2\theta} \left( \begin{array}{cc}
       c_v N_v \sin^2\theta & - c_v N_v \cos\theta\sin\theta  \\
     -  c_v N_v \cos\theta\sin\theta    & c_u N_u + c_v N_v \cos^2\theta  
    \end{array}  \right) \\
    & = \frac{1}{  c_u N_u c_v N_v \sin^2\theta} \left( \begin{array}{cc}
       c_v N_v \sin^2\theta & - c_v N_v \cos\theta\sin\theta  \\
     -  c_v N_v \cos\theta\sin\theta    & c_u N_u + c_v N_v \cos^2\theta  
    \end{array}  \right) \\
    & =  \frac{1}{  c_u N_u c_v N_v \sin^2\theta} \left[  c_u N_u     \left( \begin{array}{c}
      0  \\ 1
    \end{array} \right)
    \left(\begin{array}{cc}
      0   &  1
    \end{array} \right)
          + 
     c_v N_v \left( \begin{array}{c}
      \sin\theta  \\ -\cos\theta
    \end{array} \right)
     \left(\begin{array}{cc}
       \sin\theta  &  -\cos\theta
    \end{array} \right)
    \right] \\
    & = \frac{1}{  c_u N_u c_v N_v \sin^2\theta} 
    \left[ c_u N_u \boldsymbol{u}_\perp\boldsymbol{u}_\perp^{\ms{T}} + c_v N_v \boldsymbol{v}_\perp\boldsymbol{v}_\perp^{\ms{T}} \right].
\end{align}
Here $\boldsymbol{u}_\perp$ and $\boldsymbol{v}_\perp$ are unit vectors that are perpendicular to $\boldsymbol{u}$ and $\boldsymbol{v}$, respectively. 

According to the decomposition of the weighting vector, we want to calculate 
\begin{align}
\boldsymbol{w}^T\mathcal{F}^{-1}\boldsymbol{w} = g^2 \boldsymbol{v}^T\mathcal{F}^{-1}\boldsymbol{v}
\end{align}
Now we can easily calculate 
\begin{align}
    g^2 \boldsymbol{v}^T\mathcal{F}^{-1}\boldsymbol{v} & =    \frac{g^2 c_u N_u }{  c_u N_u c_v N_v \sin^2\theta} 
    \boldsymbol{v}^T  \left[ \boldsymbol{u}_\perp\boldsymbol{u}_\perp^{\ms{T}} \right]   \boldsymbol{v} \\
    & =  \frac{g^2 }{ c_v N_v \sin^2\theta} 
       \sqrt{N_v} \left(\begin{array}{cc}
      \cos\theta   &  \sin\theta
    \end{array} \right)   \left[   \left( \begin{array}{c}
      0  \\ 1
    \end{array} \right)
    \left(\begin{array}{cc}
      0   &  1
    \end{array} \right) \right]  \sqrt{N_v}  \left( \begin{array}{c}
      \cos\theta  \\ \sin\theta
    \end{array} \right) \\
    & =  \frac{g^2 N_v \sin^2\theta }{ c_v N_v \sin^2\theta} \\ 
    & = \frac{g^2}{c_v} \geq \frac{1}{c_v}.
\end{align}

Therefore, the optimal linear network is when $g=1$, which requires $|U_{j1}|=|U_{j2}|=\sqrt{|w_j|}$. This leads to that $\boldsymbol{w}=\boldsymbol{v}\perp \boldsymbol{u}$. A possible choice is $U_{j1}=\sqrt{|w_j|}$ and $U_{j2}=w_j/\sqrt{|w_j|}$. 

Including the cross terms, 
\begin{align}
\mathcal{F}&\approx c_u \boldsymbol{u}\boldsymbol{u}^T+c_v\boldsymbol{v}\boldsymbol{v}^T+c_s\boldsymbol{u}\boldsymbol{v}^T+c_s\boldsymbol{v}\boldsymbol{u}^T\nonumber\\
&=\left(\sqrt{c_u} \boldsymbol{u}+\frac{c_s}{\sqrt{c_u}} \boldsymbol{v}\right)\left(\sqrt{c_u} \boldsymbol{u}^T+\frac{c_s}{\sqrt{c_u}} \boldsymbol{v}^T\right)+\left(c_v-\frac{c_s^2}{c_u}\right)\boldsymbol{v}\boldsymbol{v}^T\nonumber\\
&=\tilde{c}_u\tilde{\boldsymbol{u}}\tilde{\boldsymbol{u}}^T+\tilde{c}_v\boldsymbol{v}\boldsymbol{v}^T,
\end{align}
where $\tilde{c}_u\equiv c_u-\frac{c_s^2}{c_v}$ is the effective contribution coefficient from the new vector, $\tilde{\boldsymbol{u}}\equiv \frac{1}{\sqrt{\tilde{c}_u}}\left(\sqrt{c_u} \boldsymbol{u}+\frac{c_s}{\sqrt{c_u}} \boldsymbol{v}\right)$, and $\tilde{c}_v\equiv c_v-\frac{c_s^2}{c_u}$, . According to the above analysis, $\boldsymbol{w}^T\mathcal{F}^{-1}\boldsymbol{w} \ge \boldsymbol{v}^T\mathcal{F}^{-1}\boldsymbol{v}=\frac{1}{\tilde{c}_v}$.

\section{S2. Inclusion of $\mathcal{N}$}
Including the standard quantum limit contribution $\mathcal{N}$, the inverse of the QFIM after diagonalization is approximated as
\begin{align}
\mathcal{F}^{-1}&\approx \frac{1}{\lambda_1+O(N)}\boldsymbol{e}_1\boldsymbol{e}^T_1+\frac{1}{\lambda_2+O(N)}\boldsymbol{e}_2\boldsymbol{e}^T_2+\frac{1}{O(N)}\sum_{j=1}^{2d-2}\boldsymbol{n}_j\boldsymbol{n}_j^T,
\end{align}
where $\lambda_j$ and $\boldsymbol{e}_j$ ($j=1,2$) are the eigenvalues and eigenvectors, respectively of $c_u \boldsymbol{u}^T\boldsymbol{u}+c_v\boldsymbol{v}^T\boldsymbol{v}+c_s\boldsymbol{u}\boldsymbol{v}^T+c_s\boldsymbol{v}\boldsymbol{u}^T$,
$\boldsymbol{n}_j$ is a the vector decomposed from $\mathcal{N}$ that is  orthogonal to both $\boldsymbol{e}_1$ and $\boldsymbol{e}_2$, and $O(N)$ is the contribution to each vector from $\mathcal{N}$. In the approximation of the inverse matrix, cross terms from $\boldsymbol{e}^T_j \mathcal{N} \boldsymbol{n}_j$ are omitted.

 Without loss of generality, we treat the inverse of $\mathcal{F}$ with only one vector $\boldsymbol{n}_1\equiv\boldsymbol{n}$ and the general analysis can be extended. We assume $\boldsymbol{w}=g\left(\cos\alpha \boldsymbol{u}+\sin\alpha\cos\beta \boldsymbol{v}+\sin\alpha\sin\beta \boldsymbol{n}\right)$. Evaluating the sum of the vector components on both sides leads to $0=\cos\alpha+\sin\alpha\sin\beta \left(\boldsymbol{1}\cdot\boldsymbol{n}\right)$.

Substituting the general form of the weighting vector $\boldsymbol{w}$, we obtain
\begin{align}
\label{eq:generalQFI}
\boldsymbol{w}^T\mathcal{F}^{-1}\boldsymbol{w}&\approx (\cos\alpha)^2\left(\frac{1}{\lambda_1}|g|^2 |\boldsymbol{u}\cdot \boldsymbol{e}_1|^2+\frac{1}{\lambda_2}|g|^2 |\boldsymbol{u}\cdot \boldsymbol{e}_2|^2\right)+(\sin\alpha\cos\beta )^2\left(\frac{1}{\lambda_1}|g|^2 |\boldsymbol{v}\cdot \boldsymbol{e}_1|^2+\frac{1}{\lambda_2}|g|^2 |\boldsymbol{v}\cdot \boldsymbol{e}_2|^2\right)+\frac{(\sin\alpha\sin\beta)^2}{O(N)}\nonumber\\
&=\frac{(\cos\alpha)^2}{\tilde{c}_u}+\frac{(\sin\alpha\cos\beta )^2}{\tilde{c}_v}+\frac{(\sin\alpha\sin\beta)^2}{O(N)}\nonumber\\
&=(\sin\alpha)^2\left\{\frac{1}{\tilde{c}_v}+(\sin\beta)^2\left[\frac{(\boldsymbol{1}\cdot\boldsymbol{n})^2}{\tilde{c}_u}+\frac{1}{O(N)}-\frac{1}{\tilde{c}_v}\right]\right\},
\end{align}
where we used $\cos\alpha=-\sin\alpha\sin\beta \left(\boldsymbol{1}\cdot\boldsymbol{n}\right)$ to obtain the last line. Since the second term inside the curly brackets is on the order of $\frac{1}{O(N)}$, we take $\beta=0$ to optimize the effective QFI. Accordingly, $\alpha=\pi/2$ to satisfy the symmetry of $\boldsymbol{w}$. Thus, including $\mathcal{N}$, the optimal weighting vector $\boldsymbol{w}=\boldsymbol{v}$, which also leads to $\boldsymbol{v}\perp \boldsymbol{u}$. In the limit when $\tilde{c}_v$ vanishes, i.e., a single-mode nonclassical input, we must take $\beta=\pi/2$. From Eq.~\eqref{eq:generalQFI}, $\boldsymbol{w}^T\mathcal{F}^{-1}\boldsymbol{w}$ is dominated by $\frac{1}{O(N)}$, meaning that a single-mode input state won't achieve quantum-enhanced sensitivity when the global phase involves both positive and negative weights.

Under the condition that $\boldsymbol{v}\perp \boldsymbol{u}$, we find exactly the inverse of the QFIM to be~\cite{Miller:81}
\begin{align}
\label{eq:inverseQFIM}
    \mathcal{F}^{-1}&=\frac{1}{N}\sum_j\frac{1}{4|w_j|} \ket{j}\bra{j}-\frac{c_u}{4N+c_u}\frac{2d}{4N}\ket{+}\bra{+}-\frac{c_v}{4N+c_v}\frac{2d}{4N}\ket{-}\bra{-}-\frac{2dc_s}{(4N+c_u)(4N+c_v)}\ket{+}\bra{-}\nonumber\\
    &-\frac{2dc_s}{(1+h)(4N+c_u)(4N+c_v)}\left(\ket{-}-\frac{c_s}{4N+c_u}\ket{+}\right)\left(\bra{+}-\frac{c_s}{4N+c_v}\bra{-}\right),
\end{align}
where $\ket{j}$ is the unit vector with the $j$th element equals to one, $\ket{+}=1/\sqrt{2d}(1,1,\cdots,1,1)^T$, $\ket{-}=1/\sqrt{2d}(1,-1,\cdots, 1,-1)^T$, and $h=-c_s^2/[(4N+c_u)(4N+c_v)]$.
Applying $\boldsymbol{w}=(w_1,-w_1,\cdots, w_{2d-1},-w_{2d-1})^T$ on the QFIM, we obtain
\begin{align}
    \boldsymbol{w}^T\mathcal{F}^{-1}\boldsymbol{w}=&\frac{1}{4N}\left(1-\frac{c_v}{4N+c_v}\right)-\frac{1}{4N+c_v}\frac{h}{1+h}=\frac{1}{4N+c_v-\frac{c_s^2}{4N+c_u}}.
\end{align}
Here the expressions "$4N$" appear twice in the final result due to the projections of $\mathcal{N}$ on to the vectors $\boldsymbol{u}$ and $\boldsymbol{v}$, respectively. We note that the treatment of $\mathcal{N}$ has two limitations. First, we neglected the off-diagonal terms when we expanded $\mathcal{N}$ in the orthonormal basis of $\boldsymbol{u}$ and $\boldsymbol{v}$. Second, the condition $\boldsymbol{v}\perp \boldsymbol{u}$ was optimal without the term $\mathcal{N}$. In optimizing the full QFIM, $\boldsymbol{v}\perp \boldsymbol{u}$ may not always hold.

\section{S3. $(d+1)$-mode reduced scheme}
We present an alternating protocol using a $d+1$ mode interferometer. We choose the weighting to be $\boldsymbol{w}=\left(w_1,w_2,\cdots,w_d, -\sum_{j}w_j\right)$, where the last mode serves as the reference mode for all the $d$ modes. In this case, the global quantity $ q =2\sum_{j=1}^{d+1}w_j\theta_j=2\sum_{j=1}^d w_j(\theta_j-\theta_{d+1})$. Similar to the $2d$ scheme, the optimal scheme in the $d+1$ interferometer is when the weighting vector $\boldsymbol{w}=\boldsymbol{v}$ by choosing the unitary elements of the network $U_{j1}=U_{j2}=\sqrt{w_j}$. This again leads to the condition that $\boldsymbol{v}\perp \boldsymbol{u}$. Similar to Eq.~\eqref{eq:inverseQFIM}, the inverse of the QFIM is given by
\begin{align}
    \mathcal{F}^{-1}&=\frac{1}{N}\sum_j\frac{1}{4|w_j|} \ket{j}\bra{j}-\frac{c_u}{4N+c_u}\frac{d+1}{4N}\ket{+}\bra{+}-\frac{c_v}{4N+c_v}\frac{d+1}{4N}\ket{S}\bra{S}-\frac{(d+1)c_s}{(4N+c_u)(4N+c_v)}\ket{+}\bra{S}\nonumber\\
    &-\frac{(d+1)c_s}{(1+h)(4N+c_u)(4N+c_v)}\left(\ket{S}-\frac{c_s}{4N+c_u}\ket{+}\right)\left(\bra{+}-\frac{c_s}{4N+c_v}\bra{S}\right),
\end{align}
where $\ket{S}=1/\sqrt{d+1}\text{sgn}[\boldsymbol{w}]$ and $\text{sgn}[x]$ is the sign function of $x$.
Applying $\boldsymbol{w}=\left(w_1,w_2,\cdots,w_d, -\sum_{j}w_j\right)$ on the inverse of the QFIM, we obtain the same lower bound using the condition $||\boldsymbol{w}||_1=1$
\begin{align}
    \boldsymbol{w}^T\mathcal{F}^{-1}\boldsymbol{w}=&\frac{1}{4N}\left(1-\frac{c_v}{4N+c_v}\right)-\frac{1}{4N+c_v}\frac{h}{1+h}=\frac{1}{4N+c_v-\frac{c_s^2}{4N+c_u}}.
\end{align}

\section{S4. A generic state at the first input mode}
Our protocol generalizes to a general input of $\rho_{in}=\rho\otimes\ket{\alpha}\bra{\alpha}$. The encoded state after the linear optical network $U$ and the phase encoding is given by $\rho_{en}=\exp(-i\hat{H})U\rho_{in}U^{\dagger}\exp(i\hat{H})$. A general way of evaluating the quantum Fisher information matrix is to write $\rho_{en}$ in terms of the spectral decomposition, which is equivalent to the spectral decomposition of $\rho$. Assuming the spectral decomposition of $\rho$ to be $\rho=\sum_{a}\lambda_a\ket{\lambda_a}\bra{\lambda_a}$, then the encoded state is given by 
\begin{align}
    \rho_{en}=\sum_a\lambda_a\exp(-i\hat{H})U\ket{a}\bra{a}U^{\dagger}\exp(i\hat{H}),
\end{align}
and 
\begin{align}
   \ket{a}=\ket{\lambda_a}\otimes\ket{\alpha}\otimes\ket{0}^{\otimes (2d-2)}.
\end{align}
The matrix element of the QFIM for the general input state is written as~\cite{Liu_2020}
\begin{align}
    \mathcal{F}_{jk}=2\sum_a\lambda_a\bra{a}U^{\dagger}\{\hat{n}_j,\hat{n}_k\}U\ket{a}-8\sum_{a\ne b} \frac{\lambda_a\lambda_b}{\lambda_a+\lambda_b}\Re\left[\bra{a}U^{\dagger}\hat{n}_jU\ket{b}\bra{b}U^{\dagger}\hat{n}_kU\ket{a} \right].
\end{align}
Following the appendix of Ref.~\cite{GePRL2018}, $U^{\dagger}\hat{n}_j\hat{n}_kU=\sum_{l,m,r,s}U_{jl}U_{jm}^{\ast}U_{kr}U_{ks}^{\ast} \hat{a}^{\dagger}_l\hat{a}_m\hat{a}^{\dagger}_r\hat{a}_s$ and similarly $U^{\dagger}\hat{n}_jU=\sum_{l,m}U_{jl}U_{jm}^{\ast} \hat{a}^{\dagger}_l\hat{a}_m$. Note that there are $d-2$ vacuum modes, we can evaluate the terms in $\mathcal{F}_{jk}$ based on different values of $l, m, r, s$ and obtain the following expression:
\begin{align}
    \mathcal{F}_{jk}=&|U_{j1}|^2|U_{k1}|^2\left[F_{\hat{n}_1}(\rho)-4\bar{n}_1\right]+2U_{j1}U_{j2}^{\ast}U_{k1}U_{k2}^{\ast}n_2\max_{\phi}F_{\hat{X}_{\phi}}(\rho)\nonumber\\
    &+2|U_{j1}|^2U_{k1}U_{k2}^{\ast}\Re[\mathcal{F}_{\hat{n}_1,\hat{a}_1}(\rho)\alpha^{\ast}]+2U_{j1}U_{j2}^{\ast}|U_{k1}|^2\Re[\mathcal{F}_{\hat{n}_1,\hat{a}_1}(\rho)\alpha^{\ast}]+\delta_{jk}\mathcal{N}
\end{align}
where $F_{\hat{G}}(\rho)=4\text{Tr}[\hat{G}^2\rho]-8\sum_{a\ne b}\frac{\lambda_a\lambda_b}{\lambda_a+\lambda_b}|\bra{a}\hat{G}\ket{b}|^2$ is the quantum Fisher information of a mixed state $\rho$ under the generator $\hat{G}$ with $\hat{X}_{\phi}=i(e^{-i\phi}\hat{a}_1^{\dagger}-e^{i\phi}\hat{a}_1)/\sqrt{2}$, $\mathcal{F}_{\hat{n}_1,\hat{a}_1}(\rho)=2\text{Tr}[(\hat{n}_1\hat{a}_1+\hat{a}_1\hat{n}_1)\rho]-8\sum_{a\ne b}\frac{\lambda_a\lambda_b}{\lambda_a+\lambda_b}\bra{a}\hat{n}_1\ket{b}\bra{b}\hat{a}_1\ket{a}$ is the off-diagonal element of the QFIM due to two generators $\hat{n}_1$ and $\hat{a}_1$ for $\rho$, and $\bar{n}_1=\text{Tr}[\hat{n}_1\rho]$ is the mean photon number of $\rho$. In the last term $\mathcal{N}$, $n_1$ is updated with $\bar{n}_1$ in the expression defined in the main text. Therefore, the full QFIM for the general input state $\rho_{in}=\rho\otimes\ket{\alpha}\bra{\alpha}$ has the same form as Eq.~(3) in the main text except $c_u$ is replaced with $F_{\hat{n}_1}(\rho)-4\bar{n}_1$, $c_v$ with $2n_2\max_{\phi}F_{\hat{X}_{\phi}}(\rho)$, and $c_s$ with $2\Re[\mathcal{F}_{\hat{n}_1,\hat{a}_1}(\rho)\alpha^{\ast}]$. 

One of our main results given by Eq.~(4) holds with the updated coefficients $c_u,\ c_v, \ c_s$. In the second lower bound of Eq.(4), the metrological power of a general state $\rho$ is $\mathcal{W}=\max_{\phi}F_{\hat{X}_{\phi}}(\rho)/4$~\cite{Ge:20Opearational}. Those states with $c_s=0$ saturate this lower bound, such as a squeezed thermal state. For a squeezed thermal state $\rho=\sum_{n=0}^\infty \frac{\bar{n}_{th}^n}{(\bar{n}_{th}+1)^{n+1}}\hat{S}(\xi)\ket{n}\bra{n}\hat{S}(\xi)^{\dagger}$, where $\bar{n}_{th}$ is the average thermal photon number, $\hat{S}(\xi)=\exp[(\xi^{\ast} \hat{a}_1^2-\xi\hat{a}_1^{\dagger 2})/2]$ with $\xi$ the complex squeezing number, and $\ket{n}$ is the Fock state with $n$ photon numbers. The metrological power of the squeezed thermal state is $\mathcal{W}=\left(\frac{e^{2|\xi|}}{1+2\bar{n}_{th}}-1\right)/2$~\cite{Ge:20Opearational, Kwon19} and the mean photon number of the state is $n=(2\bar{n}_{th}+1)\sinh^2 |\xi|+\bar{n}_{th}$~\cite{TanPRA:14,YuPRA:20}, which can be substituted into Eq.~(4) to evaluate the minimum estimation sensitivity using the state. In Fig. 2 of the main manuscript, we plot the minimal sensitivity scaling of a squeezed thermal state with $\bar{n}_{th}=0.1\sinh^2 |\xi|$. 

\section{S5. A four-mode distributed quantum metrology network}
In a four-mode system, the unitary elements $U_{j1}$ and $U_{j2}$ have been determined via the choice of the weighting vector. The remaining elements are $U_{j3}$ and $U_{j4}$, which should satisfy the unitary condition. 

For $w_1\cdot w_3>0$, a general set of unitary elements are given by $U_{13}=\sqrt{|w_3|}e^{i\varphi_1}$, $U_{23}=-\sqrt{|w_3|}e^{i(\varphi_1+\varphi_2-\varphi_3)}$, $U_{33}=-\sqrt{|w_1|}e^{i\varphi_1}$, $U_{43}=\sqrt{|w_1|}e^{i(\varphi_1+\varphi_2-\varphi_3)}$, $U_{14}=\sqrt{|w_3|}e^{i\varphi_3}$, $U_{24}=\sqrt{|w_3|}e^{i\varphi_2}$, $U_{34}=-\sqrt{|w_1|}e^{i\varphi_3}$, and $U_{44}=-\sqrt{|w_1|}e^{i\varphi_2}$, where the phases $\varphi_j$ are arbitrary. An example of the the unitary matrix is given in the main manuscript, where $\varphi_1=\varphi_3=0$ and $\varphi_2=\pi$.

For $w_1\cdot w_3<0$, the matrix elements are such that the values of $U_{33}\leftrightarrow U_{43}$ and $U_{34}\leftrightarrow U_{44}$.

\section{S6. Optimal measurement scheme}
We present the calculation of classical Fisher information by considering photon-number-resolving detection~\cite{Lang:13} and show that it coincides with the quantum Fisher information for a large class of input states. For the phase encoding $\sum_j\hat{n}_j\theta_j$, we insert a unitary transformation $W$ to regroup the original phases in terms of the collective phases $q_l=|\bm{w}|W_{lj}\theta_j$ via 
\begin{align}
\sum_j\hat{n}_j\theta_j=\sum_{j,k,l} \hat{n}_k W^{\ast}_{lk}W_{lj}\theta_j=\sum_{l}\hat{n}_l^{W}q_l.
\end{align}
We require $W_{1j}=w_j/|\bm{w}|$ such that $q_1=q/2$ and $\hat{n}_1^{W}=|\bm{w}|^{-2}\sum_{j}w_j\hat{n}_j$. Here $|\bm{w}|$ is the length of the vector $\bm{w}$. After the phase encoding, we make local photon-number detection with a probability of detecting $n_j^D$ in mode $j$
\begin{align}
    P(\bm {n}^D; q)=\left|\bra{\bm {n}^D}e^{i\sum_{l}\hat{n}_l^{W}q_l}U\ket{\Psi}\right|^2, 
\end{align}
where $\bra{\bm {n}^D}=\bra{n_{2d}^D}\cdots\bra{n_2^D}\bra{n_{1}^D}$ for $2d$ modes. The corresponding classical Fisher information can be obtained as
\begin{align}
F&=\sum_{\bm{n}^D}\frac{1}{P(\bm {n}^D; q)}\left(\frac{\partial P(\bm {n}^D; q)}{\partial q}\right)^2\nonumber\\
&=\sum_{\bm{n}^D}\bra{\Psi}U^{\dagger}e^{-i\sum_{l}\hat{n}_l^{W}q_l}\hat{n}_1^{W}\ket{\bm {n}^D}\bra{\bm {n}^D}e^{i\sum_{l}\hat{n}_l^{W}q_l}\hat{n}_1^{W}U\ket{\Psi}\nonumber\\
&=\bra{\Psi}U^{\dagger}(\hat{n}_1^{W})^2U\ket{\Psi}\nonumber\\
&=\frac{1}{|\bm{w}|^4}\sum_{j,k}w_jw_k\bra{\Psi}U^{\dagger}\hat{n}_j\hat{n}_kU\ket{\Psi},
\end{align}
where we assumed $\bra{\bm {n}^D}e^{i\sum_{l}\hat{n}_l^{W}q_l}U\ket{\Psi}$ is real up to a global phase in obtaining the second line in the above equation. Substituting the phase-referenced weighting condition and the unitary matrix elements, we arrive at
\begin{align}
    F=N\frac{\left(||\bm{w}||_3\right)^3}{|\bm{w}|^4}+2n_2(n_1+|\xi_1|).
\end{align}
Therefore, the estimation uncertainty of $q$ is 
\begin{align}
    \Delta q\ge \frac{1}{\sqrt{F}}=\frac{1}{\sqrt{N\frac{\left(||\bm{w}||_3\right)^3}{|\bm{w}|^4}+2n_2(n_1+|\xi_1|)}}
\end{align}
The lower bound of the above equation agrees with the lower bound in Eq. (4) of the main text under the conditions that a) $\alpha_1=\beta_1=0$ for the input state and b) $|w_j|$ are uniform for all $j$ (so that $\left(||\bm{w}||_3\right)^3=|\bm{w}|^4$). However, the second condition can be eliminated if $n_1, n_2\gg1$ such that the linear contribution of $N$ is negligible.

\end{widetext}

\end{document}